\documentclass[12pt,preprint]{aastex}
\bibpunct[, ]{(}{)}{;}{a}{}{,}

\newcommand{\thisgrb}{GRB\,050401}

\newcommand{\sect}{\S\,}

\begin{document}

\title{A $\log{N_{\rm H\,I}} = 22.6$ DLA in a dark gamma-ray burst: the environment of GRB\,050401\footnotemark[1]}


   \footnotetext[1]{Based in part on observations made at the European
                    Southern Observatory, Paranal, Chile under program
                    075.D-0270, with the Nordic Optical Telescope, operated
                    on the island of La Palma jointly by Denmark, Finland,
                    Iceland, Norway, and Sweden, in the Spanish
                    Observatorio del Roque de los Muchachos of the
                    Instituto de Astrofisica de Canarias, with the Wide
                    Field Camera (WFCAM) on the United Kingdom Infrared
                    Telescope, which is operated by the Joint Astronomy
                    Centre on behalf of the U.K. Particle Physics and
                    Astronomy Research Council, and on data collected at
                    the Subaru Telescope, which is operated by the National
                    Astronomical Observatory of Japan.}

   \author{D.~Watson,\altaffilmark{2}
           J.~P.~U.~Fynbo,\altaffilmark{2}
	   C.~Ledoux,\altaffilmark{3}
	   P.~Vreeswijk,\altaffilmark{3}
	   J.~Hjorth,\altaffilmark{2}
	   A.~Smette,\altaffilmark{3,21}
           A.~C.~Andersen,\altaffilmark{2}
           K.~Aoki,\altaffilmark{4}
	   T.~Augusteijn,\altaffilmark{5}
	   A.~P.~Beardmore,\altaffilmark{6}
	   D.~Bersier,\altaffilmark{7}
	   J.~M.~Castro~Cer\'{o}n,\altaffilmark{2}
	   P.~D'Avanzo,\altaffilmark{8,9}
	   D.~Diaz-Fraile,\altaffilmark{10}
	   J.~Gorosabel,\altaffilmark{10}
           P.~Hirst,\altaffilmark{11}
           P.~Jakobsson,\altaffilmark{2}
           B.~L.~Jensen,\altaffilmark{2}
	   N.~Kawai,\altaffilmark{12}
	   G.~Kosugi,\altaffilmark{13}
	   P.~Laursen,\altaffilmark{2}
           A.~Levan,\altaffilmark{14}
	   J.~Masegosa,\altaffilmark{10}
	   J.~N\"{a}r\"{a}nen,\altaffilmark{5}
	   K.~L.~Page,\altaffilmark{6}
	   K.~Pedersen,\altaffilmark{2}
	   A.~Pozanenko,\altaffilmark{15}
	   J.~N.~Reeves,\altaffilmark{16,17}
	   V.~Rumyantsev,\altaffilmark{18}
	   T.~Shahbaz,\altaffilmark{19}
	   D.~Sharapov,\altaffilmark{20}
           J.~Sollerman,\altaffilmark{2}
	   R.~L.~C.~Starling,\altaffilmark{22}
	   N.~Tanvir,\altaffilmark{15}
	   K.~Torstensson,\altaffilmark{5}
and	   K.~Wiersema\altaffilmark{22}}

   \altaffiltext{2}{Dark Cosmology Centre, Niels Bohr Institute, University of Copenhagen, Juliane Maries Vej 30, DK-2100 Copenhagen \O, Denmark; darach@astro.ku.dk}
   \altaffiltext{3}{European Southern Observatory, Casilla 19001, Santiago 19, Chile}
   \altaffiltext{4}{Subaru Telescope, National Astronomical Observatory of Japan, 650 North A'ohoku Place, Hilo, HI 96720, USA}
   \altaffiltext{5}{Nordic Optical Telescope, Apartado 474, Santa Cruz de La Palma, Spain}
   \altaffiltext{6}{Department of Physics and Astronomy, University of Leicester, Leicester LE1 7RH, UK}
   \altaffiltext{7}{Astrophysics Research Institute, Liverpool John Moores University, Twelve Quays House, Egerton Wharf, Birkenhead, CH41 1LD, UK}
   \altaffiltext{8}{INAF, Osservatorio Astronomico di Brera, via E. Bianchi 46, I-23807 Merate (LC), Italy}
   \altaffiltext{9}{Dipartimento di Fisica e Matematica, Universit\'{a} dell'Insubria, via Valleggio 1 1, I-22100 Como, Italy}
   \altaffiltext{10}{Instituto de Astrofisica de Andalucia, CSIC, c/ Camino Bajo de Huetor 24, 18008 Granada, Spain}
   \altaffiltext{11}{Joint Astronomy Centre, 660 North A'Ohoku Place, Hilo, HI 96720, USA}
   \altaffiltext{12}{Department of Physics, Tokyo Institute of Technology, 2-12-1 Ookayama, Meguro-ku, Tokyo 152-8551, Japan}
   \altaffiltext{13}{ALMA Project Office, National Astronomical Observatory of Japan, 2-21-1 Osawa, Mitaka, Tokyo, 181-8588, Japan}
   \altaffiltext{14}{Centre for Astrophysics Research, University of Hertfordshire, College Lane, Hatfield AL10 9AB, UK}
   \altaffiltext{15}{Space Research Institute (IKI) 117997, 84/32 Profsoyuznaya Str, Moscow, Russia}
   \altaffiltext{16}{Laboratory for High Energy Astrophysics, Code 662, NASA Goddard Space Flight Center, Greenbelt Road, Greenbelt, MD 20771, USA}
   \altaffiltext{17}{Universities Space Research Association}
   \altaffiltext{18}{Crimean Astrophysical Observatory, Ukraine}
   \altaffiltext{19}{Instituto Astrofisica de Canarias, c/ Via Lactea s/n, 38200 La Laguna, Tenerife, Spain}
   \altaffiltext{20}{Ulugh Beg Astronomical Institute, Tashkent 700052, Uzbekistan}
   \altaffiltext{21}{Research Associate, FNRS, Belgium}
   \altaffiltext{22}{Astronomical Institute `Anton Pannekoek', University of Amsterdam, Kruislaan 403, 1098 SJ Amsterdam, the Netherlands}

   \begin{abstract}
     The optical afterglow spectrum of \thisgrb\ (at $z=2.8992\pm0.0004$)
     shows the presence of a damped Ly$\alpha$ absorber (DLA), with
     $\log{N_{\rm H\,\textsc{i}}}=22.6\pm0.3$. This is the highest column density
     ever observed in a DLA, and is about five times larger than the
     strongest DLA detected so far in any QSO spectrum. From the optical
     spectrum, we also find a very large Zn column density, allowing us to
     infer an abundance of [Zn/H$] = -1.0\pm0.4$.  These large columns are
     supported by the early X-ray spectrum from \emph{Swift}-XRT which shows
     a column density (in excess of Galactic) of $\log{N_{\rm
     H}}=22.21^{+0.06}_{-0.08}$ assuming solar abundances (at $z=2.9$). The
     comparison of this X-ray column density, which is dominated by
     absorption due to $\alpha$-chain elements, and the \ion{H}{1} column
     density derived from the Ly$\alpha$ absorption line, allows us to
     derive a metallicity for the absorbing matter of
     [$\alpha$/H$]=-0.4\pm0.3$. The optical spectrum is also substantially
     reddened and can be well reproduced with a power-law with SMC
     extinction, where A$_V=0.62\pm0.06$.  But the total optical extinction
     can also be constrained in a way which is independent of the shape of
     the extinction curve: from the optical-to-X-ray spectral energy
     distribution we find, $0.5\lesssim {\rm A}_V\lesssim4.5$. However, even
     this upper limit, independent of the shape of the extinction curve, is
     still well below the dust column that is inferred from the X-ray column
     density, i.e.\ A$_V=9.1^{+1.4}_{-1.5}$. This discrepancy might be
     explained by a small dust content with high metallicity (low
     dust-to-metals ratio). `Grey' extinction cannot explain the discrepancy
     since we are comparing the metallicity to a measurement of the total
     extinction (without reference to the reddening). Little dust with high
     metallicity may be produced by sublimation of dust grains or may
     naturally exist in systems younger than a few hundred Myr. From these
     results it is clear that dust extinction properties in GRBs derived
     from comparisons of optical reddening and metallicity are unreliable.

   \end{abstract}


%
%
\section{Introduction}\label{introduction}
The largest \ion{H}{1} column density absorption line systems
[$N($\ion{H}{1}$)\geq 2\times10^{20}$\,cm$^{-2}$], known as damped
Ly$\alpha$ absorbers (DLAs), were first observed in QSO spectra, and
originate in the \ion{H}{1} regions of intervening galaxies
\citep*[see][for a recent review]{2005ARA&A..43..861W}. Because background
QSOs probe random sightlines through intervening galaxies, QSO-DLA
systems can in principle be used as an unbiased tracer of the neutral gas at
high redshift, gas that provides the fuel for star formation at these epochs
(Lanzetta, Wolfe, \& Turnshek 1995; Wolfe et~al.\ 1995)\nocite{1995ApJ...440..435L,1995ApJ...454..698W}, though complications
regarding a bias against dusty lines of sight may possibly exist
\citep{2005A&A...444..461V,2005A&A...440..499A}.

Typically, metallicities are low in QSO-DLAs 
compared to solar metallicity, ($\rm[X/H] \sim -1.5$ at $z\sim3$) with a
large scatter \citep{2002ApJ...580..732K,2003ApJ...595L...9P}.

DLAs are also common in long duration $\gamma$-ray burst (GRB) afterglow
spectra
\citep{2001A&A...370..909J,2003ApJ...597..699H,2004A&A...427..785J,2004A&A...419..927V},
typically with much higher column densities than observed in QSO-DLAs
\citep{2004A&A...419..927V}. It seems highly plausible, given that
long-duration GRBs are known to have massive star progenitors
\citep{2003ApJ...591L..17S,2003Natur.423..847H}, that the extremely large
neutral hydrogen column densities observed in many GRB-DLAs are related to
the GRBs' star-forming regions \citep*[e.g.][]{2002AJ....123.1111B}. This
means that while the highest redshift absorption system in an afterglow
spectrum belongs to the GRB host, and is therefore unlikely to be useful as
a random probe of intervening matter, it can instead provide rich
information on the sites of active star-formation in the high-redshift
universe \citep[the mean and median redshifts of recent samples of GRBs are
now very high, around $z=2.8$, and the first GRB at $z>6$ has recently been
found;][]{2006A&A...447..897J,2006Natur.440..184K,2005A&A...443L...1T,2005astro.ph..9697P,2006Natur.440..181H,2006ApJ...637L..69W}.
At the same time, afterglow spectra can still be used as random probes for
intervening absorbers at lower redshifts in the same way as QSOs. The
largest Ly$\alpha$ absorption columns in DLAs are found in GRB
afterglows: GRB\,030323 with $\log{\rm N_{H\,I}}=21.90\pm0.07$
\citep{2004A&A...419..927V} and more recently $\log{\rm
N_{H\,I}}=22.1\pm0.1$ in the afterglow of GRB\,050730
\citep{2005A&A...442L..21S,2005ApJ...634L..25C}.

Further data on the birth matrix of GRBs are provided by studies of the
afterglow's optical extinction and soft X-ray absorption. The optical
studies provide an opportunity to constrain the extinction law for
star-forming galaxies at high redshift, though the work may be complicated
by the possibility of destruction of dust grains by the GRB. Indeed it has
been proposed that GRB afterglows are subject to a `flat' or `grey'
extinction law due to a change in the grain size distribution caused by
the sublimation and breaking of dust grains
\citep*{2000ApJ...537..796W,2001ApJ...563..597F,2003ApJ...585..775P},
i.e.\ where there is evidence for little spectral reddening but where the
presence of a significant dust column is inferred. Proposed evidence for
this includes large metal column densities observed in afterglows which
exhibit little curvature in their optical afterglow spectra. The metal
column densities are derived either from optical absorption lines
\citep*[where dust depletion effects have also been
observed,][]{2003ApJ...585..638S,2004ApJ...614..293S}, or from the soft
X-ray absorption \citep{2005A&A...441...83S,2004ApJ...608..846S}, where the
most important contribution comes from oxygen K-shell absorption. This
apparent discrepancy between the metal column density and the optical
reddening was first observed in a careful and seminal paper by
\citet{2001ApJ...549L.209G}, using very low quality X-ray spectra from
\emph{BeppoSAX}.

The only alternative to a flat extinction law appears to be a non-universal
dust-to-metals ratio. This would explain a high metal column density with
little spectral reddening without reference to non-standard extinction laws,
since there is simply less dust. A low dust-to-metals ratio appears a very
natural possibility for GRB host galaxies, given their high redshifts
\citep{2006A&A...447..897J} and very young stellar populations
\citep[typically less than two hundred
Myr,][]{2003A&A...400..499L,2004A&A...425..913C,2003A&A...400..127G,2005A&A...444..711G}.
This is because metal enrichment of the interstellar medium depends on
supernovae (SNe) which have short lifetimes, while the overwhelming mass of
dust (at least in the current epoch) is not produced by SNe, but by
asymptotic giant branch (AGB) stars \citep*{2003A&A...400..981A} that have
lifetimes of at least one hundred\,Myr before even the most massive stars
begin to produce quantities of dust \citep{FerrarottiGail06}.

To distinguish between these possibilities, we require a direct measure of
the hydrogen Ly$\alpha$, soft X-ray, and optical metal absorptions in a
single GRB afterglow, observations that have been elusive because of the
conflicting requirements of a relatively low redshift for an accurate
determination of the intrinsic soft X-ray absorption, and high redshift to
move hydrogen Ly$\alpha$ to observable wavelengths.

In this paper we present X-ray and optical observations of
GRB\,050401 (\sect\ref{observations}), the resulting spectra and lightcurves
(\sect\ref{results}), and the implications of these observations
(\sect\ref{discussion}).

A cosmology with $\Omega_{\rm m} = 0.3$, $\Omega_\Lambda = 0.7$ and
$H_0=75$\,km\,s$^{-1}$\,Mpc$^{-1}$ is assumed throughout. Error ranges
quoted are 68\% confidence intervals for one parameter of interest, unless
otherwise indicated.

\clearpage
\begin{table}
 \begin{center}
 \caption{Optical and near-IR imaging observations}\label{tab:obslog}
  \begin{tabular}{@{}lllc@{}}
   \hline\hline
   Date (UT) 	& Telescope	& Mag	& Time since trigger (ks) \\[5pt]
   R-band \\
   \hline
   April 1.9880  &  MAO         & 22.7(2)       & 33.75 \\
   April 2.0720  &  TNG         & 23.0(1)       & 41.01 \\
   April 2.2100  &  VLT         & 23.27(9)      & 52.93 \\
   April 2.2243  &  NOT         & 23.0(2)       & 54.16 \\
   April 2.2563  &  VLT         & 23.31(8)      & 56.93 \\
   April 2.3004  &  VLT         & 23.44(8)      & 60.74 \\
   April 2.3183  &  D1.5        & 23.5(2)       & 62.29 \\
   April 2.3438  &  VLT         & 23.5(9)       & 64.49 \\
   April 3.2414  &  NOT         & 23.9(2)       & 142.0 \\
   April 8.2116  &  NOT         & 25.3(4)       & 571.5 \\
   April 14.5947 &  Subaru      & 26.0(1)       & 1123 \\[5pt]
   J-band \\
   \hline
   April 1.6402 & UKIRT & 18.6(2)       & 3.698 \\[5pt]
   H-band \\
   \hline
   April 1.6346 & UKIRT & 17.9(1)       & 3.214 \\
   April 1.6457 & UKIRT & 18.1(2)       & 4.173 \\
   April 1.6513 & UKIRT & 18.0(1)       & 4.657 \\
   April 2.3521 & CTIO  & 20.3(1)       & 65.21 \\[5pt]
   K-band \\
   \hline
   April 1.6242 & UKIRT & 16.88(8)      & 2.316 \\
   April 1.6284 & UKIRT & 17.2(1)       & 2.678 \\
   April 14.2742 & VLT  & $>21.5$       & 1095 \\
   \hline
  \end{tabular}
 \tablecomments{Uncertainties in the last significant digit are listed in
                parentheses after that digit.  Observations of the afterglow
                of GRB\,050401 were obtained at the ESO Very Large Telescope
                (VLT), the 8.2\,m Subaru telescope, the 2.5\,m Nordic
                Optical Telescope (NOT), the Danish 1.54\,m telescope
                (D1.5), Telescopio Nazionale Galileo (TNG), the 1.5\,m
                telescope at the Maidanak Astronomical Observatory in
                Uzbekistan (MAO), the Blanco 4.0\,m telescope at the Cerro
                Tololo Inter-American Observatory (CTIO), and the 3.8\,m UK
                Infrared Telescope (UKIRT).}
 \end{center}
\end{table}
\clearpage

%
%
\section{Observations and data reduction}\label{observations}
On 1 April 2005 at 14:20:15\,UT \emph{Swift}'s Burst Alert Telescope (BAT)
triggered on a multi-peaked burst with a duration of $33\pm2$\,s
\citep{2005GCN..3162....1B,2005GCN..3173....1S}. A rapid autonomous slew of
the satellite resulted in the X-Ray Telescope (XRT) being on source within
about two minutes and a bright, fading source was discovered in the BAT
error-circle \citep{2005GCN..3161....1A}. XRT revisited the source again on
6, 7 and 8 April.  We acquired the XRT data from the archive and used the
level~2 events files to construct the lightcurves and spectra, using a 40
pixel wide source extraction box for the Windowed Timing (WT) mode data with
a background region box 4\arcmin\ away. A 12 pixel radius circle was used to
extract the source counts in Photon Counting (PC) mode; an annulus around
the source region was used to extract a background in this case. Data in PC
mode with a source count rate above 0.4\,counts\,s$^{-1}$ are likely to be
affected by pile-up and were not used, as good statistics on the bright
afterglow were available from the WT mode data. HEASOFT\,6.0 was used to
analyse and reduce the data with the latest calibration files. Ancillary
response files were produced with \texttt{xrtmkarf} using `inarffile=CALDB',
but the fit to the data at energies close to 0.5\,keV was poor. Running
\texttt{xrtmkarf} with `inarffile=NONE' produces an ancillary response file
without using a modified on-axis ancillary reponse as input, instead using
the mirror on-axis effective area and the filter transmission file directly.
The ancillary response produced in this way improved the fit at low
energies, though the results presented here are not strongly dependent on
the choice of response file, in particular a very large excess column
density is detected regardless of which ancillary response is used.

\begin{figure*}
 \includegraphics[angle=0,width=\textwidth]{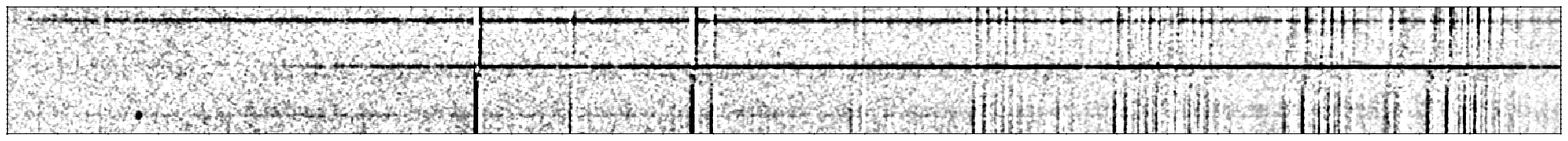}
 \vskip -0.25cm
 \includegraphics[angle=-90,width=\textwidth,bb=283 38 547 766,clip=]{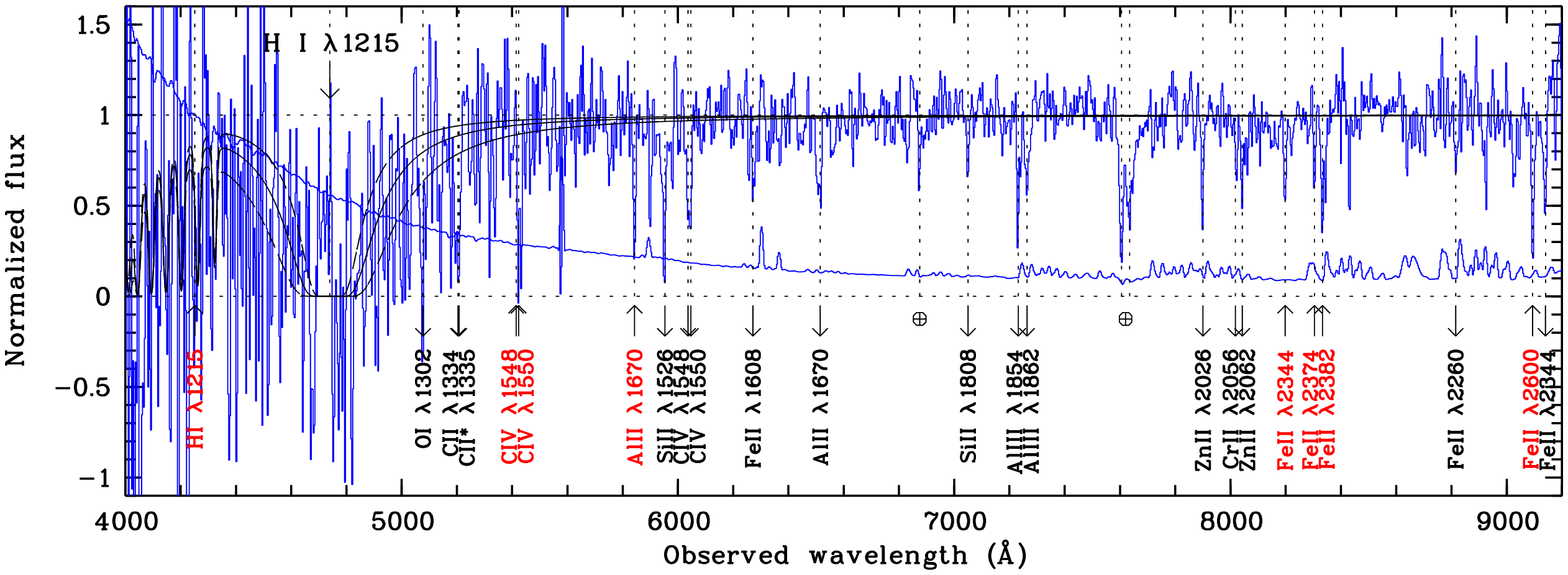}
 \caption{{\it Lower panel:} VLT/FORS2 spectrum of \thisgrb\ obtained 
          15 hours after the burst when the afterglow magnitude was
          $\mathrm{R}=23.3$.  There are two absorption systems at $z\simeq2.9$
          (black labels, down arrows) and $z\simeq2.5$ (grey labels, up arrows). 
          The DLA at $z\simeq2.9$ is the strongest ever observed and has a
          metallicity at least $\sim0.1$ times the solar value
          ([Zn/H]$=-1.0\pm0.4$).  The 1$\sigma$ error spectrum is also
          shown.  The model superimposed on the observed spectrum
          corresponds to a fitted column density of $\log N($H\,{\sc
          i}$)=22.6\pm 0.3$ together with possible H$_2$ Lyman-band lines
          from the ${\rm J}=0$ and 1 rotational levels, with $\log N({\rm
          H}_2)=21$ in both J.  {\it Upper panel:} The 2 dimensional
          spectrum covering the same wavelength range as the 1 dimensional
          spectrum in the lower panel.  The spectrum in the middle is the
          afterglow. Above and below are the spectra of 2 other objects on
          the slit. The spectrum of the object above the afterglow spectrum
          extends far bluewards of where the afterglow spectrum falls off.
          The spectrum below the afterglow spectrum is most likely due to a
          $z=2.65$ Ly$\alpha$ emitter.}
 \label{fig:ly_alpha}
\end{figure*}

Optical and near-IR observations of the optical afterglow of GRB\,050401
\citep{2005GCN..3163....1M} were secured at several telescopes (see
Table~\ref{tab:obslog}). Starting at 5:16 UT on April 2, 2005 (14.9 hours
after the burst trigger), 8 spectra of 1450\,s each were acquired with the
FOcal Reducer/low dispersion Spectrograph 2 (FORS2) instrument on Antu of the
Very Large Telescope (VLT) at Cerro Paranal in Chile, using the grism 300V and
order sorting filter GG375, and a slit width of 1\arcsec. This setting results
in a binned pixel scale of 3.24\,\AA, and a FWHM resolution of 9.5\,\AA\ at
4000\,\AA\ and 11\,\AA\ at 9000\,\AA. After cosmic ray removal using the
\citet[][]{2001PASP..113.1420V} algorithm, the spectra were reduced in the
standard fashion, using IRAF. As the night was photometric, the GRB afterglow
spectra were flux calibrated using the spectrophotometric standard star
G138$-$31.  Finally, the spectra were corrected for the Galactic foreground
extinction estimate of E(B$-$V)$ = 0.065$ mag of \citet*{1998ApJ...500..525S}.
The resulting normalized spectrum and its 1$\sigma$ error spectrum are shown in
Fig.~\ref{fig:ly_alpha}.
\clearpage
\clearpage

%
%

\section{Results}\label{results}

\subsection{The X-ray afterglow}

\begin{figure}
 \includegraphics[angle=-90,width=\columnwidth,clip=]{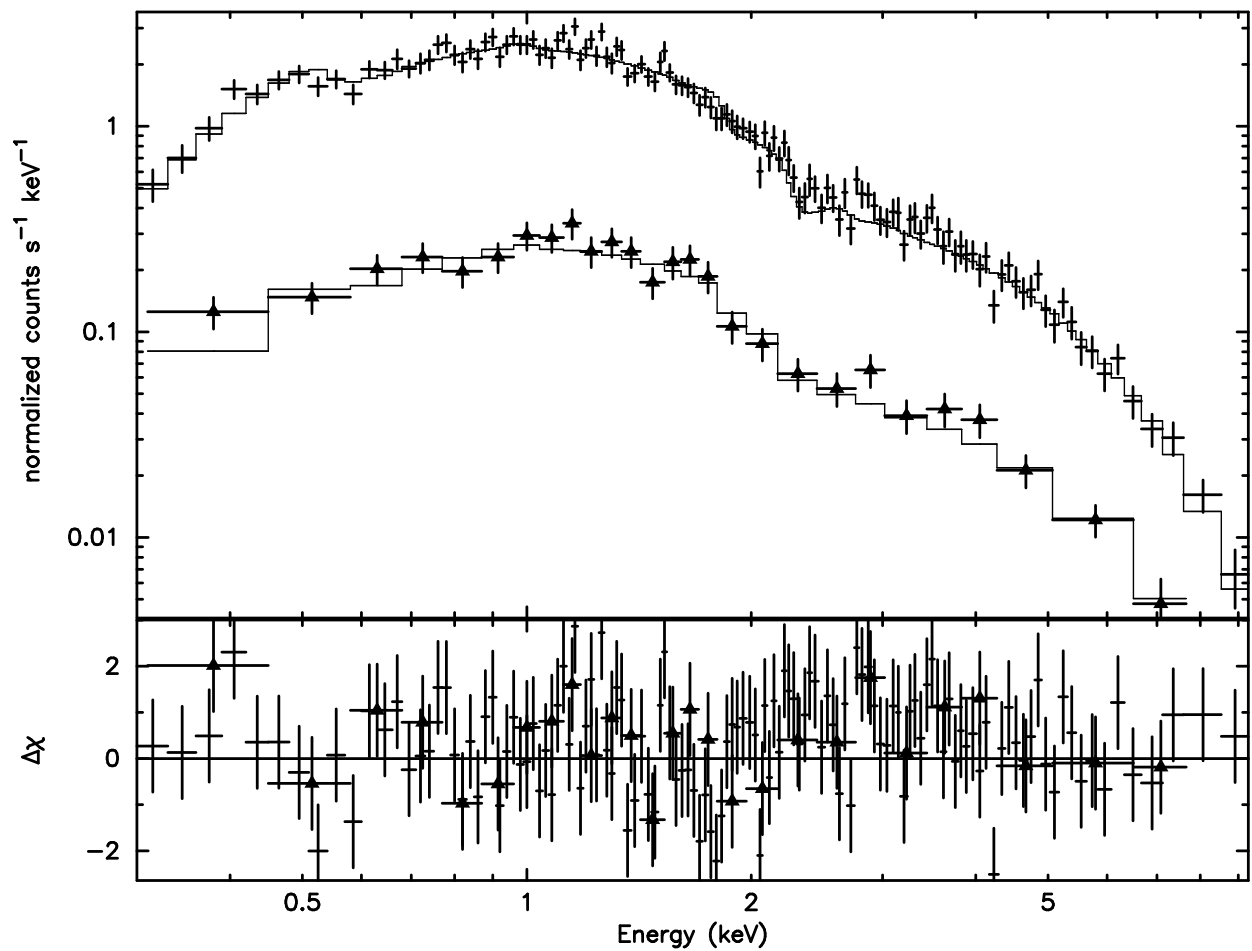}
 \caption{X-ray spectrum of the afterglow of \thisgrb\ folded through the
          instrument response (WT mode---crosses, PC mode---triangles). The
          best-fit model is plotted as a solid line. The model used is a
          power-law with absorption fixed at the Galactic level, plus
          absorption at $z=2.8992$ ($N_{\rm H}=1.6\times10^{22}$\,cm$^{-2}$
          at solar abundances).  Residuals from this model are plotted in
          the lower panel in units of standard deviations of the data.}
 \label{fig:xrtspectrum}
\end{figure}

The WT-mode spectrum (Fig.~\ref{fig:xrtspectrum}) was fit with a power-law
with fixed Galactic absorption \citep[$4.8\times10^{20}$\,cm$^{-2}$, using
the \texttt{nh} ftool,][]{1990ARA&A..28..215D}. The fit using this model was
poor ($\chi^2/{\rm dof}=629/263$), leaving large negative residuals at low
energies, and so an extra parameter, additional absorption at $z=2.8992$
(the redshift of the burst---see below), was added to the fit. This improved
the fit significantly ($\chi^2/{\rm dof}=297/262$), and gave best-fit values
of photon spectral index $\Gamma=1.89\pm0.03$ and equivalent hydrogen column
density $N_{\rm H}=1.6\pm0.1\times10^{22}$\,cm$^{-2}$ ($\log{N_{\rm
H}}=22.20\pm0.03$). Using the ancillary response file created with `xrtmkarf
inarffile=CALDB', the best-fit values were similar---$\Gamma=1.85\pm0.03$
and $N_{\rm H}=1.6\pm0.2\times10^{22}$\,cm$^{-2}$. The absorbing column
density is fit using the absorption cross-sections of
\citet{1983ApJ...270..119M}, which assumes solar abundances.
Splitting the WT-mode data in three equal duration bins revealed no
significant change in the spectrum (absorbing column or spectral slope).
Also the PC-mode data, obtained several hours after the burst, are
consistent with the same absorbing column and slope
(Fig.~\ref{fig:xrtspectrum}). An analysis of the first minute and the first
five minutes of WT-mode data revealed no variation in the absorbing column
density greater than the $1\sigma$ level. The absolute systematic
uncertainty in the flux normalisation of the spectrum is small, of the order
of a few percent.
\clearpage
\clearpage
The level of the Galactic foreground absorption may be uncertain up to
$\sim20$\% \citep{1990ARA&A..28..215D}. Including this uncertainty
leads to a more conservative error estimate of $N_{\rm
H}=1.6\pm0.3\times10^{22}$\,cm$^{-2}$ for the equivalent hydrogen column
density.

Although the X-ray column density is usually expressed in terms of the
equivalent hydrogen column density, this assumes solar abundances; the
soft X-ray absorption is in fact a direct measure of the total metal column
density in the solid and gas phases. The biggest contribution to the
absorption comes from the oxygen K-shell absorption edges, with \ion{O}{1}
at 0.52\,keV. There is also a significant contribution from other
$\alpha$-chain elements. For instance, at
$z\sim3$ the oxygen and other $\alpha$-elements are responsible for
two-thirds of the absorption in a typical X-ray CCD spectrum (e.g.\
\emph{Swift}-XRT) assuming solar abundances. 
The X-ray column density is therefore effectively a measure of the
$\alpha$-element column \citep[see
also][]{1983ApJ...270..119M,2001ApJ...549L.209G,2003ApJ...590..730T}. It is
very often impossible to determine the ionisation state of the X-ray
absorber due to 1) Galactic absorption, 2) the fact that at moderate
redshifts the oxygen edge is usually shifted out of the bandpass of most
X-ray detectors, and 3) the low resolution of CCD X-ray spectra. This,
combined with the fact that the difference in the measured absorbing column
density is expected to be small between neutral and moderate ionisation
states means that column density measures from soft X-rays are given for a
neutral column. The fact that we do not know the ionisation state of the
X-ray absorber leads, at worst, to a fairly small \emph{underestimate} of
the metal column. In this case there is evidence of a very large, neutral
column density from the optical spectrum (see below), which makes it
plausible to assume that the fraction of fully ionized gas is negligible.

\begin{figure}
 \includegraphics[angle=-90,width=\columnwidth,bb=50 61 554 734,clip=]{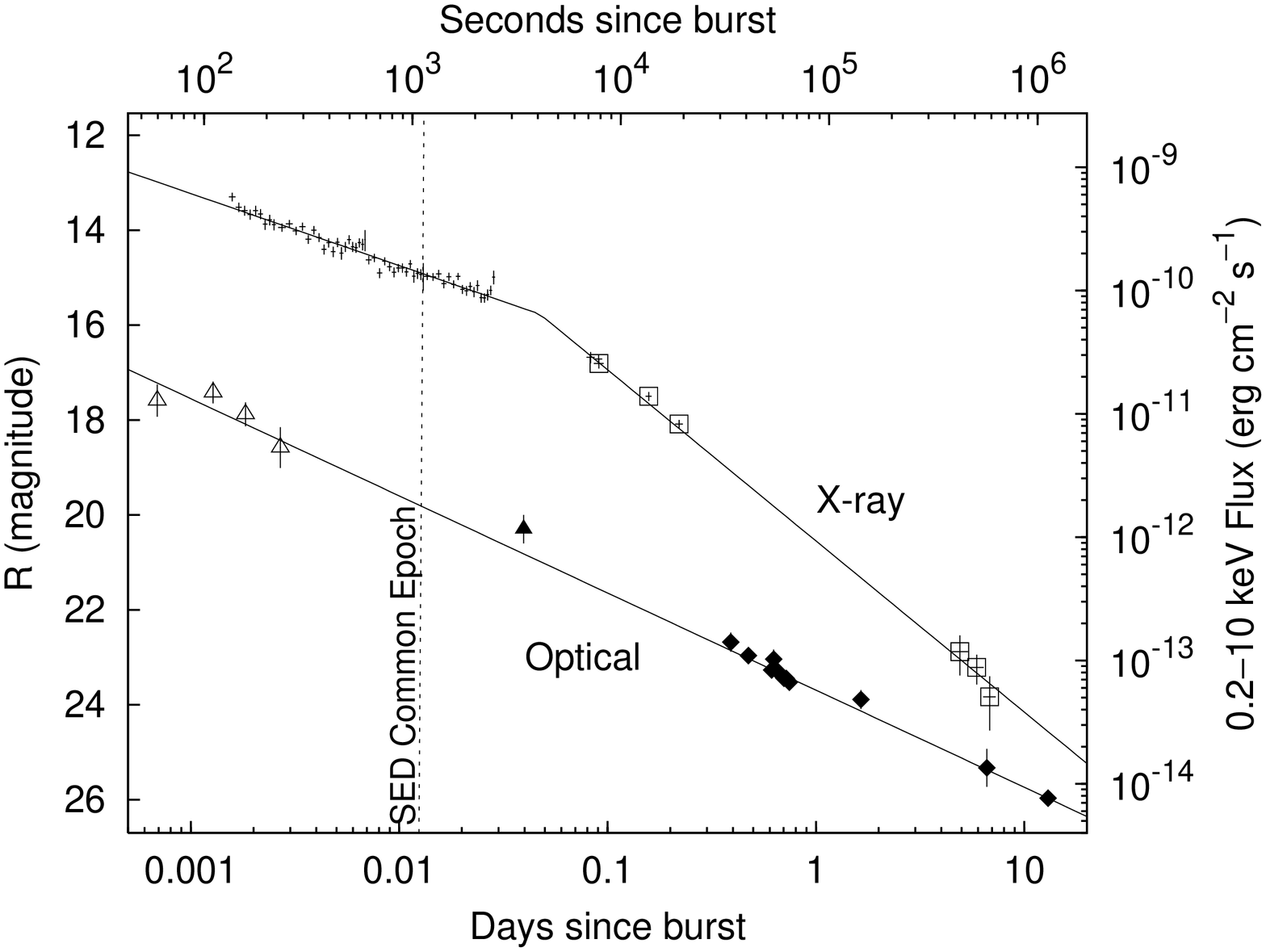}
 \caption{X-ray and optical lightcurves of the afterglow of \thisgrb.
          The R-band lightcurve (left axis, diamonds) is well fit by a
          single power-law with index $\alpha_{\rm O}=0.82$. This fit is
          consistent with data beginning at less than one minute after the
          burst, from
          \protect\citet[open triangle]{2005ApJ...631L.121R} and
          \protect\citet[filled triangle]{2005GCN..3163....1M} though they
          are not included in the fit. The observed X-ray lightcurve (right
          axis, WT mode---crosses, PC mode---boxes) requires a break at
          4090\,s with $\alpha_{\rm X}$ changing from 0.58 to 1.39. The
          epoch to which the data were interpolated to produce the SED is
          indicated with a dashed line.}
 \label{fig:lc}
\end{figure}

The 0.2--10.0\,keV lightcurve (Fig.~\ref{fig:lc}) is not well fit by a
single power-law ($\chi^2/{\rm dof}=983.5/65$) but a broken power-law
improves the fit significantly ($\chi^2/{\rm dof}=81.6/63$) with a break at
$4090\pm350$\,s, from a decay index $\alpha_{\rm X1} = 0.58\pm0.01$ to
$\alpha_{\rm X2} = 1.39^{+0.05}_{-0.04}$. There is no significant change in
the spectrum from one side of the break to the other, although the spectrum
on the late side of the break has low number counts and is not as
well-constrained as the early spectrum.
\clearpage
\clearpage
\subsection{Optical Photometry and Spectroscopy}

The optical lightcurve (Fig.~\ref{fig:lc}) is well fit with a
single power-law ($\chi^2/{\rm dof}=11.5/11$) with optical decay index
$\alpha_{\rm O}=0.82\pm0.03$. This fit is consistent with data from less than a
minute \citep[the burst was observed very rapidly by
ROTSE-IIIa,][]{2005ApJ...631L.121R} to nearly two weeks post-burst. Such a
uniform decay is unusual for optical afterglows that typically display one
or more breaks in this time interval. Furthermore, this optical decay slope
is significantly different from both X-ray decay slopes. The uniformity of
the decay, however, and the similarity with the H-band decay ($\alpha_{\rm
H}=0.76\pm0.04$) means that we can have confidence in the correction of the
optical/NIR data to an epoch common to the X-ray observations. This
confidence is important in order to reconstruct the spectral energy
distribution (SED) of the afterglow correctly.

\begin{figure*}
 \includegraphics[angle=-90,width=\textwidth]{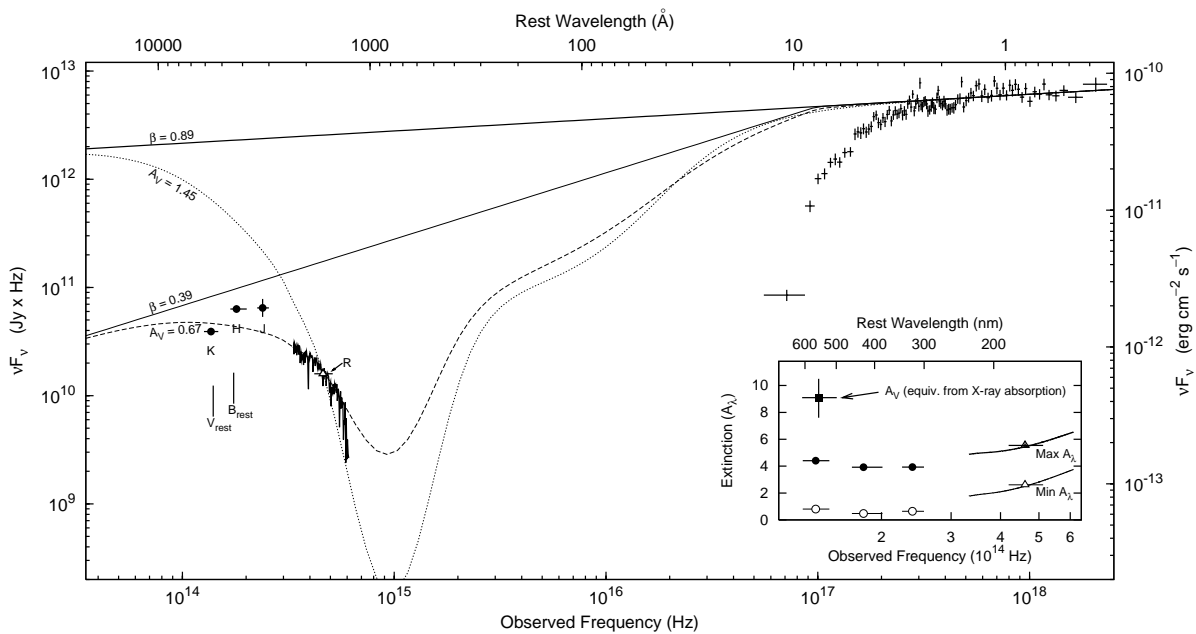}
 \caption{The spectral energy distribution of the afterglow of \thisgrb.
          The best-fit single power-law ($\beta=0.89$) to the X-ray data is
          plotted and extrapolated to optical-NIR wavelengths. All data has
          been interpolated to a common epoch of 17.9 minutes after the
          burst using the R-band lightcurve.  There is no evidence for a
          break in the X-ray power-law spectrum. To derive the minimimum
          extinction required to dim the extrapolated X-ray power-law to the
          fluxes observed in the optical-NIR, the most conservative likely
          broken power-law ($\beta=0.39$) is also plotted, with the break
          set at as high an energy as possible, at 0.4\,keV (setting the
          break at lower energies requires larger extinction, while a higher
          break energy is ruled out by the X-ray spectrum). Fitting SMC
          extinction to the optical continuum spectrum (rebinned here by a
          factor of 10) assuming an underlying $\beta=0.39$ broken
          power-law, gives a good fit with A$_V=0.67$ and is plotted as a
          dashed line.  The SMC extinction that would be required to account
          for the R-band photometric data assuming the single power-law
          ($\beta = 0.89$) continuum extrapolated from the X-rays, is
          A$_V=1.45$, and this is plotted as a dotted line for comparison. 
          However this is clearly not compatible with the observed optical
          spectral shape.
          \textit{Inset.} The maximum and minimum extinctions as a function
          of wavelength. These limits for the total extinction are derived
          from the single and broken power-laws ($\beta=0.89$ and $\beta =
          0.39$ outlined above) divided by the observed data. Circles
          represent the observed NIR data, triangles the observed R-band,
          and the solid curve the continuum from the spectral observations. 
          By contrast, the extinction expected from the soft X-ray
          absorption measurement is plotted as a filled square. The
          inconsistency of the optical/NIR limits with the extinction
          expected from the X-ray absorption shows that there is no room in
          the SED to allow for a total extinction that matches the high
          metallicity, suggesting a low dust-to-metals ratio.
          }
 \label{fig:SED}
 \label{fig:extinction}
\end{figure*}

\begin{figure}
\centerline{\includegraphics[angle=0,scale=1]{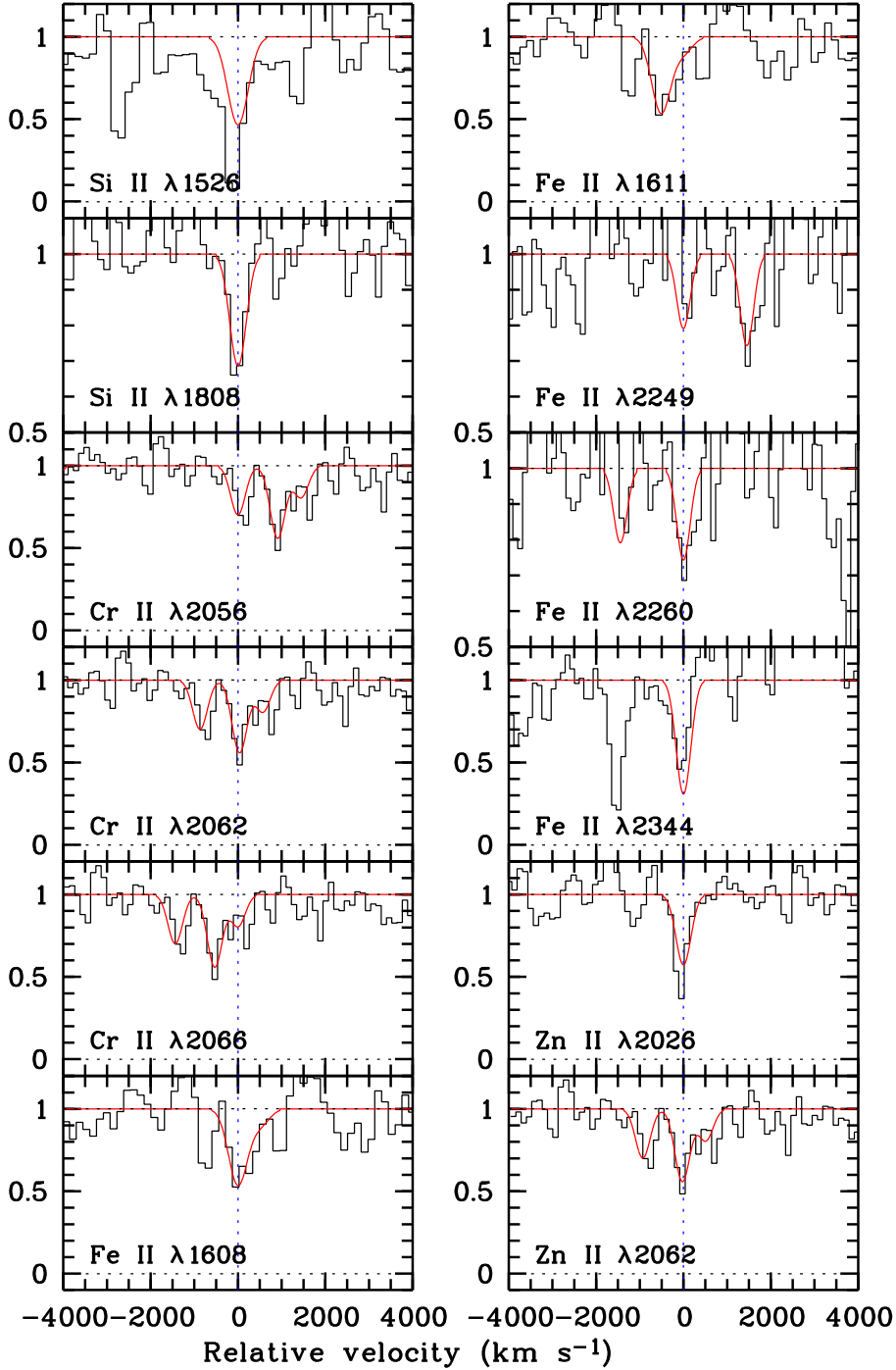}}
\vspace*{-70mm}
 \caption{Simultaneous Voigt-profile fitting of carefully selected, weaker
          and unblended lines observed at $z_{\rm abs}=2.8992$ in the
          afterglow spectrum of GRB\,050401. We also show two more
          saturated/blended lines (\ion{Si}{2} $\lambda$1526 and \ion{Fe}{2}
          $\lambda$2344) that were not used in the line fitting. The origin
          of the velocity scale corresponds to the redshift $z_{\rm
          abs}=2.8992$.}
 \label{fig:metal_lines}
\end{figure}

A striking feature in the Galactic-extinction-corrected, flux-calibrated
optical spectrum is that it is very red compared to typical optical
afterglow spectra (Fig.~\ref{fig:SED}). The best-fit spectral index
($F_\nu\propto\nu^{-\beta}$) is $\beta_{\rm O}=2.8$ (fitting the spectral
range 5800--9000\AA). This fact, coupled with an optical-to-X-ray spectral
index ($\beta_\mathrm{OX}$) of 0.2, strongly suggests that the optical
spectrum is reddened by dust. SMC, LMC and Galactic extinction laws
\citep{1992ApJ...395..130P} were fit to the optical spectrum assuming an
underlying power-law continuum with the normalisation, the underlying
power-law and the extinction as free parameters. As with all GRB afterglows
studied to date (but see Vreeswijk et al. 2005), the lack of a 2174\,\AA\
bump in the spectrum means that of these three, the SMC extinction law
provides the best fit
\citep[e.g.][]{2004A&A...419..927V,2003A&A...408..941J,2004ApJ...614..293S},
yielding an intrinsic power-law spectral index, $\beta_\mathrm{O}=0.5\pm0.2$
and $A_V=0.62\pm0.06$. We do not of course constrain the shape of the
extinction curve very well, so that this measure is heavily dependent on the
SMC curve.
\clearpage
\clearpage
We used this best-fit reddened power law and its extrapolation at wavelengths shorter
than 5800\,\AA\ to normalise the optical spectrum. A striking feature in the
spectrum is the very large absorption edge blueward of about 5400\,\AA\ (see
Fig.~\ref{fig:ly_alpha}, upper panel). Given the detection of two metal
absorption line systems in the spectrum with redshifts of $z\simeq2.5$ and
$z\simeq2.9$ (see Table~\ref{tab:alllines}) we interpret the observed very large
absorption trough as the signature of a very strong \ion{H}{1} Ly$\alpha$
absorber at $z\simeq2.9$. The correctness of this identification is based on
the observation of the red damped wing of the Ly$\alpha$ line all the way
from 4800 to 5300\,\AA\ (see Fig.~\ref{fig:ly_alpha}, lower panel). Fitting
a damped Ly$\alpha$ profile to this damped wing yields a neutral hydrogen
column density of $\log N($H\,{\sc i}$)=22.6\pm0.3$. We adopt a conservative
$1\sigma$ error of 0.3 dex which takes into account the main source of
uncertainty which is related to the continuum placement, not to the RMS
error from fitting the Voigt profile. The measurement of the rest-frame
equivalent width on the red-half side of the line (removing the
contributions of narrow absorption lines), $\sim 75$\,\AA, independently
confirms the fit result. This makes this system the largest DLA ever
observed, larger even than that observed in GRB\,050730
\citep{2005A&A...442L..21S,2005ApJ...634L..25C} and GRB\,030323
\citep{2004A&A...419..927V}, and a factor of five larger than observed in any
QSO-DLA\footnote{See
\url[http://www.ucolick.org/~xavier/SDSSDLA/tab_allfit.html]{http://www.ucolick.org/\~{}xavier/SDSSDLA/tab\_allfit.html}}.
Given the very large \ion{H}{1} column density, we checked for the presence
of H$_2$ absorption lines at the GRB redshift. In
Fig.~\ref{fig:ly_alpha} (lower panel), we display a model of the Lyman-band
lines of H$_2$ from the ${\rm J}=0$ and 1 rotational levels, assuming $\log
N({\rm H}_2)=21$ in both J. This would correspond to a molecular fraction
of $\log f\sim -1$ with $f=2N({\rm H}_2)/(2N({\rm H}_2)+N($H\,{\sc
i}$))$.  However, we cannot claim a detection given the poor quality
of the afterglow spectrum bluewards of 4500\,\AA. Among the observed metal
absorption lines, the \ion{Zn}{2} doublet at $z_{\rm abs}=2.8992$ is
unambiguously identified. The observed equivalent
width of the \ion{Zn}{2} $\lambda$2026 line is $5.5\pm 0.9$\,\AA. Since this
line may be saturated, the corresponding Zn metallicity, [Zn/H]$=-1.4$,
derived from the optically thin limit approximation, should be considered as
a lower limit. We tentatively performed simultaneous Voigt-profile fitting
of the \ion{Zn}{2} doublet lines together with carefully selected, weaker and
unblended lines from \ion{Cr}{2}, \ion{Si}{2} and \ion{Fe}{2} (see
Fig.~\ref{fig:metal_lines}). Assuming a single component and the same
(turbulent) broadening parameter value for all ions, we find a best-fit
redshift of $2.8992\pm 0.0003$ and broadening parameter
$b=67\pm35$\,km\,s$^{-1}$. The best-fit column densities are given in
Table~\ref{tab:lines}. Including the \ion{Mg}{1} $\lambda2026$ line in the
fitting process does not change the \ion{Zn}{2} column density, indicating
as observed in other GRB afterglows \citep[e.g.][]{2004ApJ...614..293S} that the
contamination of the corresponding \ion{Zn}{2} line by \ion{Mg}{1} is
negligible. With a \ion{Zn}{2} column density of
$10^{14.3\pm0.3}$\,cm$^{-2}$, \thisgrb\ also has the
highest column density \ion{Zn}{2} absorber ever recorded in a DLA
system---compare to $10^{13.95\pm0.05}$\,cm$^{-2}$ in the previous record
holder, GRB\,990123 \citep{2003ApJ...585..638S}, or
$10^{13.4\pm0.1}$\,cm$^{-2}$ in QSO\,1850+40 \citep{1998ApJ...507..113P},
the highest value so far detected in a QSO-DLA. For \thisgrb\ we find
[Zn/H]$=-1.0\pm0.4$ (taking into account the 0.3 dex uncertainty on the
determination of $\log{N({\rm H\,\textsc{i}})}$). This metallicity is
comparable to the few previous measurements for GRBs at similar redshifts,
but much higher than those of the strongest
$N$(\ion{H}{1}) GRB-DLAs observed to date
\citep[e.g.][]{2003ApJ...585..638S,2004A&A...419..927V,2004ApJ...611..200P,2005A&A...442L..21S,2005ApJ...634L..25C}.
This optically determined metallicity is lower than the metallicity obtained
from the X-ray column density. The best-fit equivalent hydrogen column
density found from the X-ray data is $\log{N_{\rm H}} =
22.21^{+0.06}_{-0.08}$ for a Solar metallicity model, which implies a
metallicity lower than Solar ([X/H$]=-0.4\pm0.3$) when compared to the
neutral hydrogen column density, $\log N($H\,{\sc i}$)=22.6\pm0.3$,
determined above. This is discussed below in \sect~\ref{discussion}.
\clearpage
\begin{table}
\begin{center}
\caption{List of all absorption lines detected with S/N$>$3}\label{tab:alllines}
\begin{tabular}{@{}lccc@{}}
\hline\hline
Observed 		& Observed	& Probable \\
Wavelength (\AA)	& Equivalent Width (\AA)	& ID\\
\hline
5074.4	& $13\pm3$	& \ion{O}{1} 1302; $z=2.8992$ \\
5178.3	&  $7\pm2$	& \\
5194.7	&  $7\pm2$	& \\[3pt]
5206.0	& $10\pm3$	& \ion{C}{2} 1334, \ion{C}{2}* 1335; $z=2.8992$ \\
5318.2	&  $6\pm2$	& \\
5422.2	&  $7\pm2$  	& \ion{C}{4} 1548; $z=2.4972$ \\[3pt]
5431.9	&  $7\pm2$  	& \ion{C}{4} 1550; $z=2.4972$ \ion{Si}{4} 1393; $z=2.8992$ \\
5841.5	& $9\pm2$ 	& \ion{Al}{2} 1670; $z=2.4972$ \\
5900.0	& $6\pm2$ 	& \\[3pt]
5950.4	& $9\pm1$ 	& \ion{Si}{2} 1526; $z=2.8992$ \\
6036.5	& $5\pm1$ 	& \ion{C}{4} 1548; $z=2.8992$ \\
6044.6	& $5\pm1$ 	& \ion{C}{4} 1550; $z=2.8992$ \\[3pt]
6273.9	& $6\pm1$ 	& \ion{Fe}{2} 1608; $z=2.8992$ \\
6559.9	& $4\pm1$ 	& \\
6507.9	& $6\pm1$ 	& \\[3pt]
6517.6	& $4.2\pm0.8$ 	& \ion{Al}{2} 1670; $z=2.8992$ \\
7049.0	& $3.2\pm0.8$ 	& \ion{Si}{2} 1808; $z=2.8992$ \\
7229.5	& $6.0\pm0.6$ 	& \ion{Al}{3} 1854; $z=2.8992$ \\[3pt]
7262.0	& $5.6\pm0.8$ 	& \ion{Al}{3} 1862; $z=2.8992$ \\
7899.2	& $5.5\pm0.9$ 	& \ion{Zn}{2} 2026; $z=2.8992$ \\
8022.6	& $3.4\pm0.9$ 	& \ion{Cr}{2} 2056; $z=2.8992$ \\[3pt]
8042.2	& $4.9\pm0.7$ 	& \ion{Zn}{2} 2062, \ion{Cr}{2} 2062; $z=2.8992$ \\
8061.6	& $2.1\pm0.7$ 	& \ion{Cr}{2} 2066; $z=2.8992$ \\
8197.9	& $5.4\pm0.7$ 	& \ion{Fe}{2} 2344; $z=2.4972$ \\[3pt]
8303.8	& $4\pm1$ 	& \ion{Fe}{2} 2374; $z=2.4972$ \\
8331.5	& $6.4\pm0.6$ 	& \ion{Fe}{2} 2382; $z=2.4972$ \\
8815.8	& $3.7\pm0.9$ 	& \ion{Fe}{2} 2260; $z=2.8992$ \\[3pt]
9023.8	& $3.7\pm0.6$	& \\
9093.7	& $9\pm1$ 	& \ion{Fe}{2} 2600; $z=2.4972$ \\
9137.6	& $6.2\pm0.8$ 	& \ion{Fe}{2} 2344; $z=2.8992$  \\
\hline
\end{tabular}
\end{center}
\end{table}

\begin{table}
\begin{center}
\caption{Absorption line fits at $z=2.8992$}\label{tab:lines}
\begin{tabular}{@{}lcccc@{}}
\hline\hline
Central (\AA) & $\log{N}$	& ID	& [X/H]	& [X/Zn]\\
Wavelength	& (cm$^{-2}$)	& 	& 	&	\\
\hline
7049.8(6)  & 16.5(4)  & \ion{Si}{2} 1808  & $-1.7$(5)  & $-0.7$(5)\\
7900.3(7)  & 14.3(3)  & \ion{Zn}{2} 2026  & $-1.0$(4)  & \\
8042.7(7)  & 14.3(3)  & \ion{Zn}{2} 2062  & 	       & \\
6271.7(5)  & 16.0(2)  & \ion{Fe}{2} 1608  & $-2.1$(4)  & $-1.1$(4)\\
6282.4(5)  & 16.0(2)  & \ion{Fe}{2} 1611  &            & \\
8772.7(8)  & 16.0(2)  & \ion{Fe}{2} 2249  &            & \\
8815.2(8)  & 16.0(2)  & \ion{Fe}{2} 2260  &            & \\
8017.8(7)  & 14.6(2)  & \ion{Cr}{2} 2056  & $-1.7$(4)  & $-0.7$(4)\\
8041.1(7)  & 14.6(2)  & \ion{Cr}{2} 2062  & 	       & \\
8056.4(7)  & 14.6(2)  & \ion{Cr}{2} 2066  &            & \\
  \hline
 \end{tabular}
 \tablecomments{The uncertainty in the last significant digit is
                listed in parentheses after that digit.}
 \end{center}
\end{table}

\clearpage
\clearpage
We note that there is an intervening absorption system at $z=2.4972$ based on
\ion{Si}{2}, \ion{Al}{2}, and \ion{Fe}{2} lines. The metal lines in this 
system are strong but significantly weaker than those from the GRB system.
Moreover, the \ion{Zn}{2} lines from this system are not detected. It is not
possible to estimate the \ion{H}{1} column density from the Ly$\alpha$ line
as there is no flux left from the afterglow around 4255\,\AA. Assuming a
metallicity of a tenth solar we estimate $\log N($H\,{\sc i}$)\sim21$. This
intervening DLA therefore provides a negligible contribution to the
uncertainties on the absorbing column in X-rays or the dust extinction.
 
In the 2 dimensional spectrum we also identify a Ly$\alpha$ emitter at
$z=2.65$ at an impact parameter of 10.2\arcsec\ south of the GRB position
(81.3\,kpc at $z=2.65$, see the upper panel in Fig.~\ref{fig:ly_alpha}). 
There is no significant absorption in the afterglow spectrum at this
redshift. There is also no evidence for Ly$\alpha$ emission from the host
galaxy in the 2D spectrum, but as the host galaxy is not detected this does
not exclude even a large equivalent width Ly$\alpha$ emission line from the
host.

\subsection{The spectral energy distribution}

The photon-weighted mean time of the WT-mode X-ray spectrum was 1075\,s
after the burst. The flux of the optical spectrum was interpolated to this
time, using the $\alpha_{\rm O}=0.82$ optical decay slope and the flux
calibration of the spectrum. The best-fit to the R-band lightcurve is
consistent with the flux interpolation of the optical spectrum to this time
(Fig.~\ref{fig:SED}). The NIR data were also corrected to this time,
assuming their decay was the same as the R-band decay rate. The correction
in the NIR was less than for the optical spectrum since the NIR observations
occurred between 37 and 76 minutes after the burst.

The SED (Fig.~\ref{fig:SED}) clearly shows that a break in the
optical-to-X-ray spectral index is required even to fit a simple SMC
extinction-correction to the optical spectrum alone. Larger absorption with
less reddening would of course allow the optical spectrum to be fit if we
assume an intrinsic optical slope that was a continuation of the X-ray
slope. Such an extinction would be consistent with the idea of a flat or
`grey' extinction as advocated by \citet{2001ApJ...549L.209G},
\citet{2004ApJ...614..293S} and \citet{2005A&A...441...83S} among others.

Limits on the total extinction (A$_\lambda$) for \thisgrb\ were derived from
the extreme cases of 1) a single power-law extrapolated from the X-ray
spectrum to the optical ($\beta=0.89$) and 2) a broken power-law with a
$\Delta\beta=-0.5$ (giving $\beta=0.39$), as expected from the fireball
model \citep*{1998ApJ...497L..17S}, with the break set conservatively just
inside the observed X-ray band (0.4\,keV). These limits on A$_\lambda$ are
shown in the inset to Fig~\ref{fig:extinction}. It is notable that even a
very flat extinction curve cannot alone reconcile the upper limit to the
extinction with the large metal column densities we observe. We discuss this
conflict below.

%
%
\section{Discussion}\label{discussion}

The optical afterglow of \thisgrb\ is particularly faint---in fact it is the
faintest GRB afterglow for which an absorption line redshift has been
obtained. The observed optical to X-ray spectral slope is flatter than
expected for the fireball model, therefore \thisgrb\ falls into the `dark
burst' population as defined by \citet{2004ApJ...617L..21J}. The observed
reddening in the optical spectrum allow us to conclude that the `dark'
nature of the burst can be explained by extinction in the host galaxy. It is
interesting that the very high column densities observed in GRB hosts, which
are probably local to the GRBs, means that we do not need to invoke very
dusty environments to explain dark bursts---even with a small dust-to-gas
ratio these very high column densities are sufficient to darken the
afterglows.

\subsection{Metallicity}
Since we expect GRB progenitors to be short-lived
\citep{2003Natur.423..847H,2003ApJ...591L..17S}, the cloud probed by the
afterglow is likely to be the matrix out of which the progenitor was born. For
this reason, it may be problematic to try to infer the global properties of the
ISM at high-redshift from the GRB's probable host DLA.  But for exactly the
same reason GRB-DLAs are particularly important for determining the nature of
the progenitor material of the GRB. For instance, the favoured model of long
duration GRBs \citep[the
`collapsar',][]{1993ApJ...405..273W,2006ApJ...637..914W}, predicts that GRBs
should generally be found in low metallicity environments.  Alternative
progenitor models exist where there is no such low metallicity preference 
\citep[e.g.][]{2005ApJ...632.1001O,2005ApJ...623..302F}.

Previous observations of GRB afterglows and GRB host galaxies have indeed
suggested that GRBs reside in low metallicity galaxies
\citep{2003A&A...400..499L,2003A&A...406L..63F,2004ApJ...611..200P}, and in
fact the two next strongest GRB-DLAs after GRB\,050401 have rather low
metallicities
\citep{2004A&A...419..927V,2005A&A...442L..21S,2005ApJ...634L..25C}. The
absorbers in GRB afterglows usually have high metal column densities, but
relatively little reddening---this phenomenon has been attributed to a flat
extinction law, often suggested to be due to destruction of dust grains
\citep{2001ApJ...549L.209G,2001ApJ...563..597F,2004ApJ...614..293S,2005A&A...441...83S}.

Here we have one of the first direct measures in a single GRB of the
\ion{H}{1} column density and the total metal (primarily $\alpha$-element)
column density. If we assume that the X-ray and optical absorptions probe
the same matter, and there is no evidence to the contrary, this allows us to
constrain the metallicity from the X-ray measurement, suggesting
[X/H]$=-0.4\pm0.3$, where most of the uncertainty is in the hydrogen column.
Using the best-fit to the optical absorption lines we find
[Zn/H]$=-1.0\pm0.4$. Zn is one of the elements least affected by depletion
onto dust grains, at least in warm clouds in the Galaxy
\citep{1996ARA&A..34..279S}, and is also a good tracer of the Fe-group
metallicity. By fixing the hydrogen column density at the level of the
optically-determined \ion{H}{1} column and the Fe-group elements at
the level determined from the Zn abundance, we can fit for the
$\alpha$-element abundance in the X-ray spectrum. We obtain
$\alpha=0.36\pm0.06\,Z_\odot$, which translates to [$\alpha$/H]$=-0.4\pm0.3$,
including the \ion{H}{1} uncertainty. We can also directly obtain the ratio of
$\alpha$-chain metallicity (from the X-ray spectrum) to Fe-group metallicity
(from the optical spectrum). This ratio may hint at a possible overabundance of
$\alpha$-elements to Zn compared to solar: [$\alpha$/Zn]$=0.6\pm0.3$ (a
factor of about 4--5).

In an analysis of QSO absorption lines systems, \citet{2003ApJ...590..730T}
invoke an overabundance of $\alpha$-chain to Fe-group elements of 2.5. Such
an overabundance is consistent with the results obtained here. Indeed, this
degree of $\alpha$-element overabundance is consistent with the Zn
metallicity \citep[see][]{2003ApJ...590..730T,2005PASA...22...49M}, and
would indicate a relatively young ISM still dominated by products from core
collapse rather than type~Ia supernovae \citep{2005PASA...22...49M}.

From the optical measurements, there is no evidence for an overabundance of
Si, which is also considered an $\alpha$-chain element, compared to Zn, with
[Si/Zn]$=-0.7\pm0.5$. Since Si is depleted onto dust grains as discussed
above, this ratio probably reflects differential dust depletion effects.
Evidence for dust depletion comes from iron for which we measure
[Fe/Zn]$=-1.1\pm0.4$ similar to the highest seen in QSO-DLAs
\citep{2003MNRAS.346..209L}.


In cold Galactic clouds Zn is depleted by about $-0.7$ dex
\citep{1996ARA&A..34..279S}, so it is important to examine if the apparent
alpha-element overabundance could be due to a large depletion of Zn onto
dust grains. We believe the depletion of Zn is not large in this case, as
the abundances of Si, Cr, and Fe relative to Zn in this system
([Si/Zn]$=-0.7\pm0.5$, [Cr/Zn]$=-0.7\pm0.4$, [Fe/Zn]$=-1.1\pm0.4$) are more
similar to the element ratios in warm Galactic clouds ([Si/Zn]$\sim-0.5$,
[Cr/Zn]$\sim-1.0$, [Fe/Zn]$\sim-1.2$) than those in cold Galactic clouds
([Si/Zn]$\sim-0.6$, [Cr/Zn]$\sim-1.6$, [Fe/Zn]$\sim-1.6$) \citep[see][for
details]{1996ARA&A..34..279S}. Nevertheless, we cannot exclude some
depletion of Zn onto dust and in this case the total metallicity would be
higher if metal ratios were similar to the solar values
([X/H]$=-0.5\pm0.3$). Strictly, the derived $\alpha$-element overabundance
should hence be considered an upper limit.

\subsection{Extinction}
The optical spectrum is well fit using an SMC extinction curve with
A$_V=0.62\pm0.06$ and an intrinsic slope (before dust extinction) of $\beta
= 0.5\pm0.2$. With a slightly flatter slope of $\beta=0.39$, as expected
from the X-ray slope $\beta_X=0.89$ and a cooling break of $\Delta \beta =
0.5$ between the X-rays and the optical, the required extinction is slightly
higher, A$_V=0.67$ (see Fig.~\ref{fig:extinction}).

This reddening-derived estimate is somewhat less than might be expected from
a simple translation of the optical \ion{H}{1} or \ion{Zn}{2} columns to
extinction. These would respectively suggest A$_V=2.3^{+1.9}_{-1.2}$
assuming an SMC dust-to-hydrogen ratio \citep{1985A&A...149..330B}, or
A$_V=2.2^{+2.3}_{-1.1}$, following \citet{2004ApJ...614..293S}. The
similarity of these latter two estimates is not surprising of course since
the estimated metallicity in the SMC is similar to the [Zn/H] observed here.
However, these estimates are strongly dependent on the metallicity and if
some Zn is depleted onto dust, the metallicity-estimated A$_V$ could be
substantially greater.

A more robust, direct comparison can be made between the optical spectrum
and the optical-to-X-ray SED, thereby placing bounds on the total and
selective extinctions (see the inset in Fig.~\ref{fig:extinction}). 
A$_V<0.5$ is excluded: it would require a non-power-law continuum in the
optical which was also substantially below the continuum expected from the
X-ray spectrum, even with a steep spectral break. On the other hand, we can
also exclude A$_V>4.5$ since it requires not only an extremely flat
extinction curve, but also that the optical and X-ray afterglows are driven
by different emission components (e.g. that the X-ray emission is dominated
by inverse compton radiation rather than synchrotron radiation).

The absolute limits on the extinction ($4.5>{\rm A}_V>0.5$) and the
extinction estimates based on metallicity and on reddening-fitting examined
above, are all much less than the A$_V$ inferred from the metal column
derived from the soft X-ray absorption, which is $9.1_{-1.5}^{+1.4}$
\citep{1995A&A...293..889P}. Since the X-ray absorption is dominated by
metals, the conversion from X-ray absorption to dust column is not sensitive
to a gas-to-metals conversion, but to the dust-to-metals conversion. 
Allowing an $\alpha$-element overabundance as suggested above would not make
these more compatible since it is the $\alpha$-elements that compose most
the dust. 
We are therefore forced to invoke a dust-to-metals ratio (as distinct from
the dust-to-gas ratio or the metallicity) which is smaller in this DLA than
in the Galaxy or the Magellanic clouds. If the reddening-derived extinction
measure is correct, this is more than a factor of ten less than the SMC
dust-to-metals ratio. But such a resolution to the low-reddening vs.\ high
metal column problem clashes with the apparent pattern of dust depletion
observed above, as depletion at this level naturally implies the presence of
a large dust column along with the gas. We are left with a paradox: we very
clearly observe a large metal column density and insufficient total
extinction (and reddening) implying a low dust-to-metals ratio, but at the
same time we have some evidence of dust depletion effects in the ratios of
the optical lines. This is puzzling. We cannot exclude the hypothesis that
there is a large dust column which has very little effect on the
transmission of optical light. On the other hand it seems reasonable to say
that the evidence for dust depletion (4--5$\sigma$ in total) may not be
quite strong enough to support such a far-reaching conclusion. In that case
we invoke a low dust-to-metals ratio as above. Such a low dust content with
a moderately high metallicity would indicate either sublimation of most of
the dust along the line of sight \citep[e.g.][]{2002ApJ...569..780D}, or
more simply, that the dust observed in this DLA system has been formed by
SNe rather than AGB stars. This is because SNe contribute metals and dust
very quickly to the ISM, but while they provide nearly all of the metals,
they produce only relatively small amounts of dust \citep[SNe produce at
most only a few percent of the dust found in the local
universe;][]{FerrarottiGail06,2005ASPC..341..539N}.

\section{Conclusions}

We have reported the strongest damped Ly$\alpha$ line observed to date, in
the spectrum of GRB\,050401. The metallicity is fairly high even for a
GRB-DLA, with [Zn/H$]=-1.0\pm0.4$.  Large X-ray absorption is also detected,
from which a large $\alpha$-element column density is inferred. The
extinction expected based on the X-ray absorbing column density exceeds the
total extinction allowed by the SED in GRB\,050401, and is more than an
order of magnitude greater than the extinction inferred from a good fit of
the SMC extinction curve to the reddening in the optical spectrum. A grey or
flat extinction law cannot explain this discrepancy. We suggest that since
the metallicity is high, and the total extinction is low, it implies that
there is comparatively little dust along the line of sight to this GRB
(however there is puzzling evidence to the contrary in the depletion
patterns of the optical absorption lines). This high metal column with
little extinction may be evidence of dust destruction, or that the stellar
population is very young and that little dust has been formed yet, in
spite of the vigorous star-formation activity. Based on these results,
inferences on dust extinction properties drawn from comparisons of optical
reddening and metallicity should be approached cautiously.

\acknowledgments
We thank Peter Capak for excellent support of our CTIO observations. The
Dark Cosmology Centre is funded by the DNRF. We acknowledge benefits from
collaboration within the EU FP5 Research Training Network, `Gamma-Ray
Bursts: An Enigma and a Tool'.

\par We note that \citet{2006MNRAS.365.1031D} have published an analysis of
the afterglow of GRB\,050401 while this paper was in review. The analysis of
the XRT data by \citeauthor{2006MNRAS.365.1031D} produced compatible results
with those obtained here. The optical afterglow lightcurve published here is
more complete and yields a steeper single power-law decay. We also show here
that grey dust is not a sufficient explanation for the discrepancy between
the apparent extinction in the optical/UV and the large X-ray column.

\end{document}